\documentclass[12pt,showpacs,amsmath,showkeys,preprint,letter]{revtex4}
\usepackage{graphicx}
\usepackage{color}
\usepackage{amssymb}
\usepackage[latin9]{inputenc}

\begin{document}

\title{Anderson Localization Phenomenon in One-dimensional Elastic Systems}


\author{R.~A.~\surname{M\'endez-S\'anchez}}
\affiliation{Instituto de Ciencias F\'isicas, Universidad Nacional
Aut\'onoma de M\'exico, P.O. Box 48-3, 62251 Cuernavaca, Mor., Mexico}

\author{J.~Flores}
\affiliation{Instituto de F\'{\i}sica, Universidad Nacional
Aut\'onoma de M\'exico, P.O. Box 20-364, 01000 M\'exico, D. F.,
Mexico}

\author{A.~D{\'{\i}}az-de-Anda}
\affiliation{Instituto de F\'{\i}sica, Universidad Nacional
Aut\'onoma de M\'exico, P.O. Box 20-364, 01000 M\'exico, D. F.,
Mexico}

\author{L.~Guti\'errez}
\affiliation{Instituto de Ciencias F\'isicas, Universidad Nacional
Aut\'onoma de M\'exico, P.O. Box 48-3, 62251 Cuernavaca, Mor.,
Mexico}

\author{G.~Monsivais}
\affiliation{Instituto de F\'{\i}sica, Universidad Nacional
Aut\'onoma de M\'exico, P.O. Box 20-364, 01000 M\'exico, D. F., Mexico}


\author{A.~Morales}
\affiliation{Instituto de Ciencias F\'isicas, Universidad Nacional
Aut\'onoma de M\'exico, P.O. Box 48-3, 62251 Cuernavaca, Mor., Mexico}

\begin{abstract}

The phenomenon of Anderson localization of waves in elastic systems
is studied. We analyze this phenomenon in two different set of
systems: disordered linear chains of harmonic oscillators and
disordered rods which oscillate with torsional waves. The first set
is analyzed numerically whereas the second one is studied both
experimentally and theoretically. In particular, we discuss the
localization properties of the waves as a function of the frequency.
In doing that we have used the inverse participation ratio, which is
related to the localization length. We find that the normal modes
localize exponentially according to Anderson theory. In the elastic
systems, the localization length decreases with frequency. This
behavior is in contrast with what happens in analogous quantum
mechanical systems, for which the localization length grows with
energy. This difference is explained by means of the properties of
the reflection coefficient of a single scatterer in each case.
\end{abstract}



\maketitle

\section{\label{Intro}Introduction}

The discovery of Anderson localization phenomenon in
quantum mechanics gave origin to one of the most important subjects
in condensed matter physics since it has a crucial effect on the
transport properties of materials. As a matter of fact, the original
work of Anderson~\cite{Anderson} is among the most cited papers of
the twentieth century. The theory of Anderson
localization~\cite{Anderson,KramerMacKinnon} studies the alterations
in the localization properties of the wave functions brought about
by disorder in the system. It is well known that in a perfect
periodic material the allowed energy levels form a band structure
and the wave functions associated with the allowed energies are
extended along the whole system. In this case, when an electric
field is applied to the material and the energy of the electrons is
such that there exist empty levels with energy close to the Fermi
energy, the electrons can move throughout the material and an
electrical current is produced. However, if the system has random
imperfections, for example when there are strange atoms in an
otherwise ideal chemical composition or when there is abnormal
spacing between some atoms due to dislocations, the wave functions
could be localized in some region of the system, therefore affecting
the conductivity. In the particular case of one-dimensional infinite
disordered systems, any amount of disorder produces localization in
all the wave functions except for a set of states with zero measure.
Thus, band theory and the theory of Anderson localization allow us
to understand why some materials conduct electricity and why others
do not.

Anderson localization has also been observed in many classical wave
like phenomena: in
optics~\cite{dalichaouch,Chabanov,StorzerGrossAegerterMaret,aegerter,SchwartzBartalFishmanSegev,lagini},
elasticity~\cite{HeMaynard,HeMaynard2,Maynard,HuStrybulevychPageSkipetrovVanTiggelen,floresanderson},
water waves~\cite{lindelof} and cold atomic gases~\cite{garreau}. 
When the systems are translational invariant, on the one hand, the elastic wave amplitudes are extended. One also finds a band spectrum of frequencies in this case. The wave amplitudes become localized in the disordered elastic systems, on the other hand, as is also true in solid state physics.
However, in the elastic experiments one can go beyond what is
obtained in quantum mechanics, since we can measure the wave
amplitudes at each point. This allows us to understand the Anderson
localization phenomenon in a deeper way. In
this work we study this phenomenon in two different sets of elastic
systems: the special linear harmonic oscillator chains described in
Fig.~\ref{Fig_resorte} of Section~\ref{Sec:Chains}, and the
disordered elastic rods defined in Fig.~\ref{Fig_Varilla_Desord} of
Section~\ref{Sec:Rods}. For the first set of systems we present
numerical results while for the disordered rods this work has an
experimental part and a theoretical one. To discuss the wave
localization we calculate the inverse participation ratio. We also
present a brief discussion of the mean free path $\ell_m$.

\section{\label{Sec:Chains}Linear Harmonic Oscillator Chain}

\begin{figure}
  \includegraphics[width=\columnwidth]{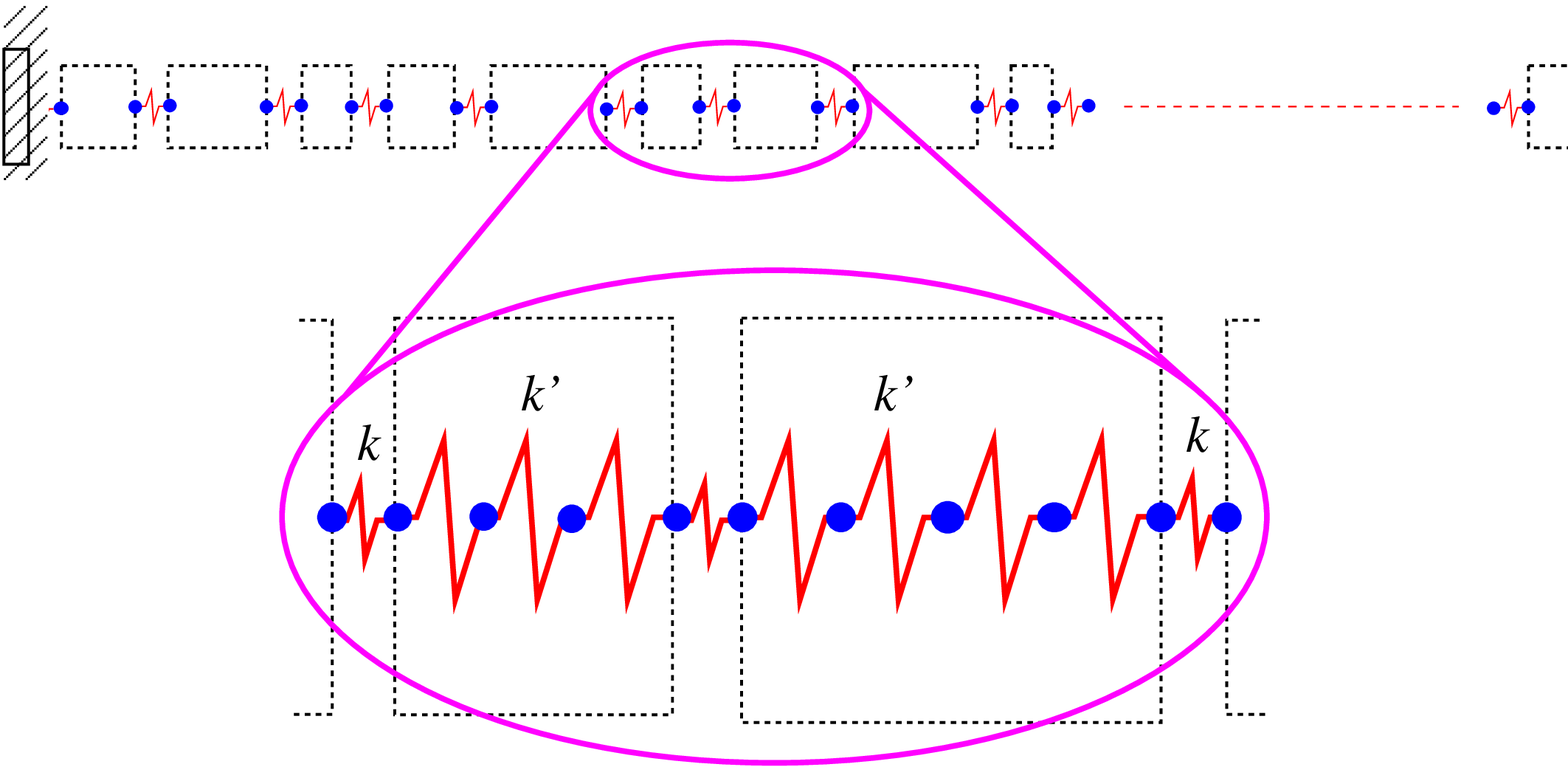}
  \caption{Disordered chain of harmonic oscillators.
  The zoom shows an amplification of two particular blocks
  with $N_6=3$ and $N_7=4$.}
  \label{Fig_resorte}
\end{figure}

In this Section we analyze the first ensemble of elastic systems.
Each member of the ensemble is a linear harmonic oscillator chain
(LHOC), formed by $n$ identical point particles of mass $m$
coupled by $(n+1)$ springs, which in the general case have different
strengths $k_{i}$. The two ends of each LHOC are fixed to walls with
infinite mass. For homogeneous systems all the springs are equal to
each other and the normal-mode frequencies can be obtained in closed
form, as first done by Lagrange in 1788 \cite{Lagrange1788}. In the
general case, the problem can be solved using the following
well-known formalism.

The Lagrangian of the linear chain is
\begin{equation}
L=\dfrac{m}{2}\sum_{j=1}^{n}{\dot\xi_{j}^{2}-\frac{1}{2}\sum_{j=1}^{n+1}k_{j}\left(\xi_{j}-\xi_{j-1}\right)^{2}}
\end{equation}
where $\xi_{j} (\textit{t})$ is the value of the coordinate of the
\textit{j}-th mass measured with respect its equilibrium value
$\xi_{j}^{0}$; therefore $\xi_{n+1}=\xi_{0}=0,$ since they
correspond to the infinite mass walls. The Lagrange equations take
the form
\begin{equation}
m\ddot{\xi_{j}}=-k_{j}\xi_{j-1}+\left(k_{j}+k_{j+1}\right)\xi_{j}-k_{j+1}\xi_{j+1}.
\end{equation}
When all the masses move with the normal mode of frequency
$\omega_{I}$, that is, when
$\xi_{j}\left(t\right)=\psi_{j}^{I}\exp\left(-i\omega_{I}t\right)$,
where $\psi_{j}^{I}$ is the oscillation amplitude of mass \textit{j}
in the normal mode \textit{I}, the following eigenvalue equation is
obtained:
\begin{equation}
-k_{j}\psi_{j-1}^{I}+\left[\left(k_{j}+k_{j+1}\right)-m\omega_{I}^{2}
\right]\psi_{j}^{I}-k_{j+1}\psi_{j+1}^{I}=0.
\label{Eq:Ecus}
\end{equation}
Solving Eq.~(\ref{Eq:Ecus}) is equivalent to diagonalizing the
matrix \textbf{H} with elements
\begin{eqnarray}
H_{j,j-1}&=&-k_{j}\nonumber\\
H_{j,j}&=&k_{j}+k_{j+1}\label{Eq:Hij}\\
H_{j,j+1}&=&-k_{j+1}\nonumber
\end{eqnarray}
and $H_{i,j}=0$ otherwise. In this process one obtains the
normal-mode frequencies $\omega_{I}$ and the corresponding
eigenvectors
$\Psi_{I}=\left(\psi_{1}^{I},\psi_{2}^{I},\ldots,\psi_{n}^{I}\right)$.

Many different sets $\left\{ k_{i}\right\} $ have been analyzed in
the past~\cite{Ashcroft}. Here, we shall introduce a special set
which behaves exactly as a vibrating rod with notches in
compressional oscillations. We introduce a set of $\cal N$ blocks,
each consisting of $N_{j}$ springs with strength $k'$ and connected
to their adjacent neighboring blocks by a spring of constant $k$.
See Fig.~\ref{Fig_resorte}. When $k'>>k$ this system of blocks
behave, for compressional oscillations, in a similar way as a
vibrating elastic rod with notches; the soft springs play the role
of the notches. Indeed, as will be published elsewhere, when all
$N_{j}$ are identical to each other, the oscillator chain produces
the band spectra we have found for a locally periodic rod with
notches \cite{Moralesetal}. Furthermore, we have also found that
changing $N_{j}$ in such a way that
\begin{equation}
 N_j=\left[{\frac{N}{1+j\gamma}}\right],
\end{equation}
where $[x]$ means the largest integer not greater than $x$ and $\gamma$ is
a real number, the Wannier-Stark ladder we encountered in~\cite{Gutierrezetal} is regained. Our oscillator chains are, as a
consequence, useful to study disordered elastic systems and to learn
how Anderson localization emerges.

We form the ensemble of LHOC's in the following way: the family of 
numbers $\left\{ N_{j}\right\} $, is a set of  uncorrelated integer
random numbers. The ensemble is defined with a uniform distribution
in the interval
$\left[N\left(1-\Delta\right),N\left(1+\Delta\right)\right]$. Here
$N$ is the average of $N_{j}$ and $\Delta$ measures the disorder.
In the results we show here, we used $N=70$ and $\Delta=2/7$.
Each member of the ensemble has in general a different number of
masses and springs.

\begin{figure}
  \includegraphics[width=0.5\columnwidth]{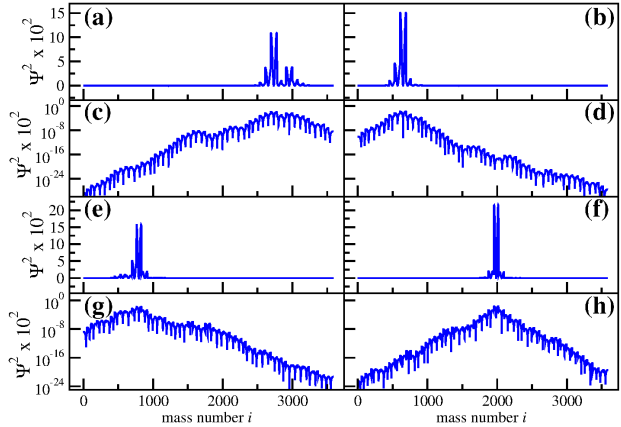}
  \caption{Square of wave amplitudes of four particular normal modes of a disordered linear chain.}
  \label{Fig_AmpCadena}
\end{figure}

We shall now present some results that show that Anderson
localization arises. In Fig.~\ref{Fig_AmpCadena}, we show examples
of $\left(\psi_{j}^{I}\right)^{2}$ 
for some particular normal-mode frequencies $\omega_{I}$ of a given member of the
ensemble with ${\cal N}=50$. In Figs.~\ref{Fig_AmpCadena}(a), (b), (e), and (f), the values of
$\left(\psi_{j}^{I}\right)^{2}$ are plotted in arbitrary units,
whereas in Figs.~\ref{Fig_AmpCadena}(c), (d), (g), and (h), the same wave amplitudes 
are given but in a semi-log scale, respectively. The four values of
\textit{I} considered in this figure correspond to wave functions
with 49, 54, 72, and 82 nodes, respectively. From these figures the
localization of wave amplitudes is evident. Since the logarithmic
plots are straight lines at both sides of the maxima, the wave
amplitudes decrease exponentially on the average. One should mention
that the wave functions with a number of nodes much less than $\cal N$ 
are not appreciably altered by the disorder since the wavelength
is larger than the block size.

To discuss the localization properties of the ensemble of oscillator
chains, we have calculated the inverse participation ratio (IPR).
For a given eigenstate $\Psi^{I}$ the IPR is given by
$\sum_{j}\left[\psi^{I}_j\right]^{4}$ and measures the number of
sites that contribute significantly to the eigenfunction
normalization. For states with an exponential decay, such as those
given in Fig.~\ref{Fig_AmpCadena}, the IPR is directly connected with the localization
length $\xi$~\cite{Fyodorov}.

\begin{figure}
\includegraphics[width=0.5\columnwidth]{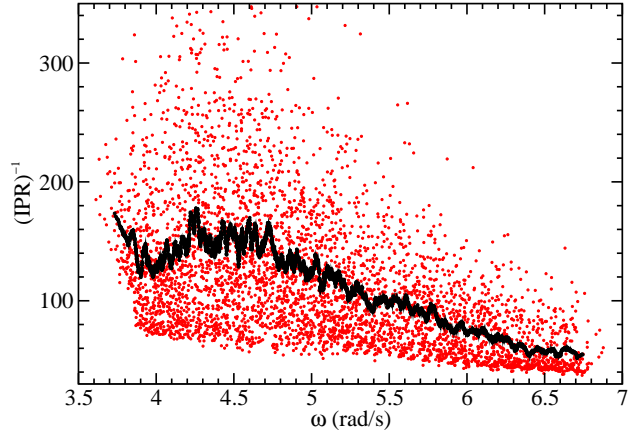}
\caption{IPR$^{-1}$ as a function of $\omega$.
The points are obtained from 50 different
disordered chains with ${\cal N}=50$, $N=70$, $\Delta=2/7$ and $n\approx 3500$. 
The continuous line is a window average.}
  \label{Fig_IPRvsW}
\end{figure}

To calculate the average of the IPR we have considered an ensemble
of systems like the one defined in Fig.~\ref{Fig_resorte}. Then, we
have calculated the IPR for all the eigenfunctions of the ensemble
which have a normal-mode frequency in a certain eigenvalue interval.
The points of Fig.~\ref{Fig_IPRvsW} represent the values of the
IPR$^{-1}$ of 50 different disordered chains. 
The continuous line shows the window average.

\begin{figure}
  \includegraphics[width=0.5\columnwidth]{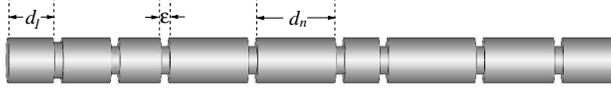}
  \caption{Disordered rod which consists in a set of sub-rods
  of different length. The values of the parameters are $R_L=1.28\,{\rm cm}$,
  $\eta=R_{\varepsilon}/R_L=0.65$, $d=7.2\,{\rm cm}$, $\Delta=0.35$,
  $\varepsilon=1.016\,{\rm mm}$ and $N=50$.}
  \label{Fig_Varilla_Desord}
\end{figure}

\begin{figure}
\includegraphics[width=0.5\columnwidth]{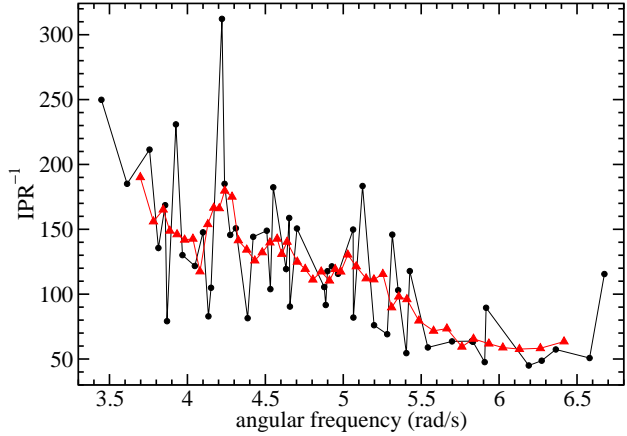}
\caption{IPR$^{-1}$ as a function of $\omega$ for
the special LHOC that simulates the real rod studied in the
laboratory. The red triangles indicate a window average of 20
eigenvalues.} 
\label{Fig_IPR_osc_Varilla}
\end{figure}

In order to construct the LHOC that simulates the behavior of a
particular disordered rod we use the following procedure. Let us
consider the rod of Fig.~\ref{Fig_Varilla_Desord}. If the length of
the $j-th$ sub-rod is $d_j~$milimeters, then one constructs a block
consisting of a linear chain with $[d_j]$ masses. Here $[d_j]$ means
the largest integer not greater than $d_j$. These masses are joined by $[d_j]-1$
springs of strength $k^{\prime}$.  Finally, as mentioned before,
each block of masses is coupled with its neighboring blocks by means
of springs of strength $k\ll k^{\prime}$. We have built, also
with this procedure, the equivalent LHOC associated with the real rod
studied in the laboratory. In Fig.~\ref{Fig_IPR_osc_Varilla} we show
the values of the IPR$^{-1}$ as a function of frequency $\omega$ for
this LHOC (full circles).  We observe that the IPR$^{-1}$ decreases
with frequency. The continuous line indicates a window average.

Measuring the time displacements of many particles subject to different springs is, however, a tough problem, which can become very cumbersome. Therefore, to perform experiments the oscillator chain will be replaced by an elastic rod with notches. 
We have already mentioned that the compressional modes for these rods behave identically to the special LHOC used here. However, we shall consider in the next section results for torsional vibrations for the two following reasons: 
On the one hand, one expects that compressional and torsional modes behave in a similar way since they obey the same equation; on the other hand, torsional waves are easier to measure with the experimental setup that we use.
%
%
As we will see below, the  IPR$^{-1}$ of Fig.~\ref{Fig_IPR_osc_Varilla} shows indeed similar results to those measured for the torsional modes of the rod given in Fig.~\ref{Fig_IPRvarilla}. 


\section{\label{Sec:Rods}Inhomogeneous Elastic Rod}

We now study the torsional waves in rods with notches
spaced at random along their axis. Each rod consists of a set of
$\cal N$ coupled sub-rods with the same axis and radius $R_L$ and
whose lengths are $d_1, d_2,...,d_{\cal N}$. The family $\{d_i\}$ is
a set of $\cal N$ random uncorrelated lengths with a uniform
distribution in the interval $[d(1-\Delta),d(1+\Delta)]$. Here $d$
is the average of $d_i$ and $\Delta$ measures the disorder. As shown
in Fig.~\ref{Fig_Varilla_Desord} the $\cal N$ sub-rods are joined by
identical short cylinders of radius $R_{\varepsilon}=\eta R_L$ and
length $\varepsilon \ll d_i$, $\forall i$. The parameter $\eta$, the
coupling constant, satisfies the relation $0<\eta\le 1$. For the
experiment we have analyzed only one rod which was machined from a
single aluminum piece. It should be mentioned that in the numerical
simulations we have considered an ideal case of free conditions at
the ends of the rod which is an approximation of the real
configuration since the rod is supported by means of two thin
threads. The frequency range we used in the experiment was 0-87~kHz
and therefore
\begin{equation}
R_L=1.28 {\rm \,cm}<\lambda_{\rm min}=\frac{c_T}{f_{\rm max}}=\frac{3140 {\rm \,m/s}}
{87000 {\rm \,Hz}}=3.60 {\rm \,cm},
\label{Eq_Rl}
\end{equation}
which means that the cross section of the rod is not excited, so it
behaves as a quasi one-dimensional system. The value of $c_T$ was
measured by fitting the spectrum of the aluminium rod before
machining the notches.

We have calculated numerically the set of eigenvalues $\{f_i\}$ and
the corresponding eigenfunctions, using the Poincar\'e map
method~\cite{AvilaMendez-Sanchez}. Then, the IPR as a function of
frequency was obtained and an average of the localization length in
some intervals of frequencies was calculated. The calculations were
done with an effective value of the parameter $\eta$  as discussed
in Refs.~\cite{Moralesetal,Moralesgutierrezflores}.

\begin{figure}
\includegraphics[width=0.5\columnwidth]{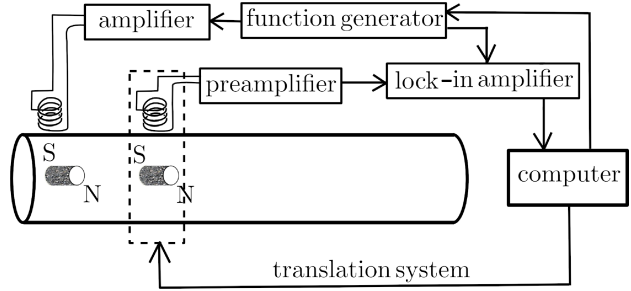}
\caption{The experimental setup used to measure normal-mode
amplitudes and frequencies of the rod of
Fig.~\ref{Fig_Varilla_Desord}. Torsional waves are excited with the
EMAT on the left side of this figure, whereas the detection is made
with the EMAT placed on the translation system. In both EMAT's the
coil axis is perpendicular to the permanent magnet axis and these in
turn are orthogonal to the rod axis.} 
\label{Fig:ExperimentalSetup}
\end{figure}

To perform the measurements, the electromagnetic-acoustic transducer
(EMAT) developed by us~\cite{Moralesetal}, was used. The EMAT
consists of a permanent magnet and a coil, and can be used either to
detect or excite the oscillations. The transducer operates through
the interaction of eddy currents in the metallic rod with a
permanent magnetic field. According to the relative position of the
magnet and the coil, the EMAT can either excite or detect
selectively compressional, torsional or flexural vibrations. Used as
a detector, the EMAT measures acceleration. This transducer has the
advantage of operating without mechanical contact with the rod. This
is crucial to avoid perturbing the shape of the localized wave
amplitudes. Both the detector and exciter are moved automatically
along the rod axis by a computerized control system so the wave
amplitudes can be measured easily (see
Fig.~\ref{Fig:ExperimentalSetup}).

Whenever the wave amplitude is measured, it is necessary to keep the
system at resonance, so the wave generator must be as stable as
possible. In order to achieve this, a Stanford Research Systems DS345 wave generator with an ovenized time base with stability $<0.01$
ppm was used in the experiment. This is enough when variations of
the bar temperature are not important. In our case this condition
was satisfied since the measurements were made in a very short time
interval.

\begin{figure}
\includegraphics[width=0.5\columnwidth]{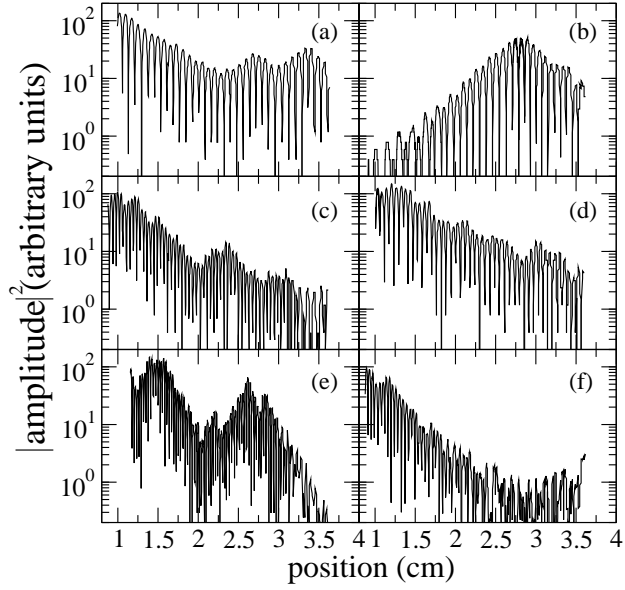}
\caption{Measured localized wave amplitudes: (a)~$f=18220$~Hz,
(b)~$f=18942$~Hz, (c)~$f=34370$~Hz, (d)~$f=25380$~Hz, (e)~
$f=51553$~Hz, and (f)~$f=35560$~Hz.} \label{Fig.WA1}
\end{figure}

Before presenting the experimental and numerical results, we shall
consider what we will call an independent rod model. This provides a
qualitative argument to understand why all normal modes of a
disordered rod are localized. The small sub-rods of length $d_i$ are
independent of each other when $\eta\to 0$. The $i-{\rm th}$ sub-rod
is excited when the driving force has a frequency $f$ equal to
$f_i^{(q)}=pc_{T}/2d_i$, where $c_{T}$ is the speed of torsional
waves and $p$ is an integer. The other sub-rods are, generally, not
excited since $d_j$ is usually different from $d_i$. Hence, the
amplitude of the vibration decreases and the wave amplitude is
localized. Several experimental examples of this fact are shown in
Fig.~(\ref{Fig.WA1}b), (\ref{Fig.WA1}d) and (\ref{Fig.WA1}f), where
we have plotted the logarithm of the wave amplitude as a function of
the coordinate $x$ along the rod. We observe that the envelope of
the logarithm is a straight line at both sides of the maximum, which
implies that the wave amplitude decreases as an exponential
function. However, it could also happen that the length of some
other sub-rod, say $d_k$, be almost equal to $d_i$. The amplitude of
the vibration could then present two maxima. In
Figs.~(\ref{Fig.WA1}a), (\ref{Fig.WA1}c) and (\ref{Fig.WA1}e) this
case is apparent; we observe that they also decay exponentially.
When the disorder is very small, that is $\Delta \ll1$, the above
argument implies that all the sub-rods can be excited with a driving
frequency $f\sim pc_{T}/2d$  so the localization of the wave
amplitude grows and it could even exceed the total length $L$ of the
complete rod.

From this independent rod model we see that introducing disorder in
the set $\{d_i\}$, whose first effect is to produce a disordered set
of normal-mode frequencies, is a way to simulate a diagonal disorder
in a quantum one-dimensional tight-binding hamiltonian where the
coupling $\eta$ between first neighbors is equal to a constant.
Therefore, the rods considered here are quasi one-dimensional
disordered systems where the frequency plays the role of the energy
in quantum mechanics. If all the sub-rods of radius $R_L$ have the
same length, the vibrations of the rod are extended waves traveling
in a periodic structure.

\begin{figure}
\includegraphics[width=0.5\columnwidth]{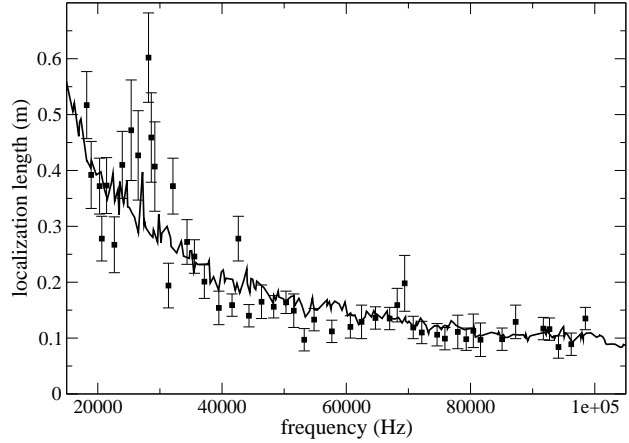}
\caption{The localization length $\xi$ as a function of the frequency $f$.
  The dots correspond to the experimental measurements and the continuous line
  to the numerical average using an ensemble of 3000 rods.}
  \label{Fig_8}
\end{figure}

\begin{figure}
\includegraphics[width=0.5\columnwidth]{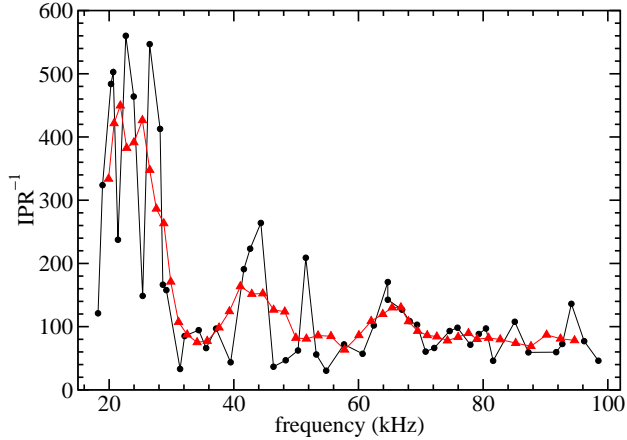}
\caption{IPR$^{-1}$ as a function of the frequency $f$ for the rod studied in the laboratory. The triangles indicate a window average of 20 eigenvalues.}
  \label{Fig_IPRvarilla}
\end{figure}

In the laboratory we have proceeded as follows. We have measured the
values of the amplitude of each eigenfunction at a sufficiently
large number of points of the rod and then, by using a least-squares
fit, we have matched a function of the form $C\,\exp(-|x-x^*|/\xi)$ to their envelope; here $x^*$ is the position of
the maximum of the amplitude. This provides us with the localization
length $\xi$. Several eigenfunctions in each frequency interval were
considered and the average of $\xi$ was evaluated. In most cases, we
were able to measure the right-hand side of the wave amplitude only,
because the exciter, placed at the left-hand side of the rod,
produces a magnetic field in this zone that affects the measurements
of the detector. With the purpose of adopting a systematic rule to
define $\xi$  and its experimental error, we have taken into account
only the side of the wave amplitude that decays exponentially
towards the increasing values of $x$. The fit was always made taking
into account only 10 points.

For some eigenmodes the fit could be done in two or three zones, so
several values of $\xi$ were obtained. In these cases $\xi$ was
defined as the average of these values. In Fig.~\ref{Fig_8} we show the
experimental and numerical average values of $\xi$ as a function of
the frequency $f$. 
A good agreement between the experiment and the
numerical simulations is obtained. It is found that, in the elastic
rods, the localization length $\xi$ of the normal modes decreases
with the frequency. This is corroborated obtaining the
participation ratio for the experimental rod; the IPR$^{-1}$ is
shown in Fig.~\ref{Fig_IPRvarilla}.

\begin{figure}
 \includegraphics[width=0.5\columnwidth]{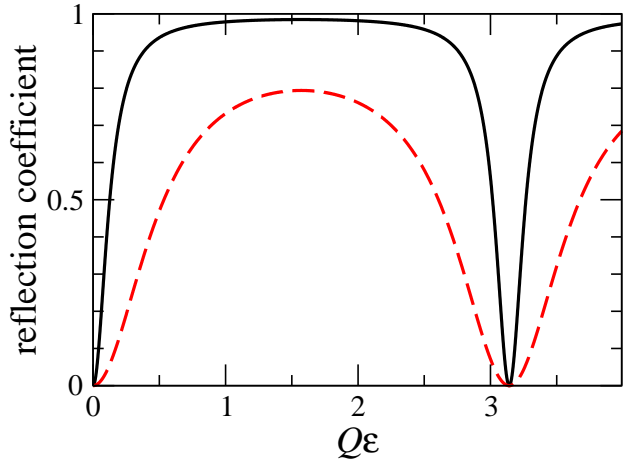}
 \caption{Reflection coefficient $|r|^2_{\rm classical}$ as a function of $Q \varepsilon$ for  torsional waves through a notch. The continuous and dashed lines correspond to $\eta=0.5$ and $\eta=0.7$, respectively.}
 \label{Fig_ReflectionCoefficient}
\end{figure}

One should note that in the quantum mechanical analog, $\xi$ grows
with the energy eigenvalue. This difference is not surprising if one
analyzes the behavior of the individual scatterers that constitute
the system in each case: a notch for the rod and a potential barrier
for the quantum case. In particular, when one calculates the  reflection coefficient $|r|^2$ as a function of the eigenvalue
for the case of a notch one obtains:
\begin{equation}
|r|^2_{\rm classical}={4{\rm sen}^2(Q\varepsilon)\over
    \Big({\eta^4-1\over \eta^4+1}-{\eta^4+1\over \eta^4-1}\Big)^2+
    4{\rm sen}^2 (Q\varepsilon)},
\label{Eq_singlenotch}
\end{equation}
whereas for the case of a quantum particle of mass $m$ in
the presence of a rectangular barrier of height $U_0$ and width
$\varepsilon$ one gets
\begin{equation}
|r|^2_{\rm quantum}={4{\rm sen}^2(q\varepsilon)\over
    \Big({\tilde{\eta}^4-1\over \tilde{\eta}^4+1}-{\tilde{\eta}^4+1\over \tilde{\eta}^4-1}\Big)^2+
    4{\rm sen}^2 (q\varepsilon)},
\label{Eq_single_barrier}
\end{equation}
where $Q={2\pi f\over c_T}$, $q=\sqrt{{2m\over
{\hbar}^2}(E-U_0)}$, $\tilde{\eta}=\big({E-U_0\over E}\big)^{1/8}$,
and $\hbar$ is Planck's constant divided by $2\pi$. 
In Fig.~\ref{Fig_ReflectionCoefficient} we show a plot of $|r|^2_{\rm classical}$ as a function of $Q\varepsilon$. We can see that this is an increasing function of the eigenvalue in the range of frequencies we use, whereas in the quantum case it decreases with the energy. 
The crucial difference between the two cases is that in the elastic case $\eta$ is a
constant determined by the geometry of the system, whereas in the quantum case 
$\tilde\eta$ is an increasing function of the energy. For these
simple scatterers, we have also calculated the mean free path
$\ell_m$ defined as $\ell_{m}\equiv\big(|r|^2/\Delta x\big)^{-1}$,
where $|r|^2/\Delta x$ is the reflection coefficient per unit
length. In our case we have taken $\Delta x=\varepsilon$. As a
consequence of the previous discussion, the behavior of $\ell_{m}$
in the classical system we have considered is opposite to that in
the quantum case. In particular, the classical $\ell_{m}$ is a
decreasing function of the frequency whereas for the quantum case
it is an increasing function of the energy.

Note that in the classical case a reflection coefficient that
increases with frequency indicates that the smaller the wavelength
is, the more noticeable the defects, {\it i.e.,} the notches, are.
Therefore the defects tend to decrease the transmission when the
wave length also decreases. In the quantum case one has a reflection
coefficient that decreases with energy which indicates that the
greater the wavelength is, the defects, {\it i.e.},  the potential
barriers, are less important. This fundamental difference between
the quantum model and the classical system reflects the incapacity
to simulate quantum potentials using rods with notches.
Consequently, it is expected that in our experiments the dependence
of the localization length as a function of frequency be opposite to
that observed in the quantum case. Nevertheless, the Anderson
localization phenomenon appears in both situations.


We now return to the discussion of systems with a large number of
scatterers. Since the waves show an exponential decrease, it is
expected that the square of the transmission coefficient
$|t|^2=1-|r|^2$ shows an exponential decrease of the form
\begin{equation}
|t|^2=\exp\Big(-{x\over {\ell}^{\prime}_m}\Big),
\label{Eq_t2}
\end{equation}
where $\ell^{\prime}_m$ is a constant which, as is easy to
see, is equal to the mean free path $\ell_m$. In fact, if we
evaluate the above expression for  $x=\Delta x \ll \ell^{\prime}_m$ we
have
\begin{equation}
|t|^2=1-|r|^2=\exp\Big(-{\Delta x\over {\ell}^{\prime}_m}\Big)=
    1-{\Delta x\over \ell^{\prime}_m}
\label{Eq_t2Delta}
\end{equation}
which implies that $\ell_m=\ell^{\prime}_m$.

\section{\label{Sec:Conclu}Conclusions}

In the present work two different sets of elastic
disordered structures were analyzed. We have presented, both numerically
and experimentally, evidence of the Anderson localization. The localization length of the normal modes as a function of frequency was measured on elastic rods by means of the
wave amplitude as well as by the inverse participation ratio, and not by means of the transmittance. A good agreement between the numerical simulations and the experimental
values was obtained. We have found that the localization length and
the participation ratio decrease with the frequency, which is
opposite to the quantum mechanical case. 

\section{acknowledgements}

This work is supported by DGAPA-UNAM under project IN113011. 
RAMS was supported by DGAPA-UNAM under project IN111311 and by CONACYT under project 79613. GM was supported by DGAPA UNAM project IN-117712-3. We are indebted to Pierric Mora for his help in the measurements.

\end{document}